\documentclass[a4paper,11pt]{article}

\usepackage{epsfig}
\usepackage{amssymb}

\newcommand{\be}{\begin{equation}}
\newcommand{\ee}{\end{equation}}
\def\e{{\rm e}}
\newcommand{\erf}{\rm{erf}}

\def\k{\kappa}

\title{A note on the Voigt profile function\footnote
{Submitted to: {\it J. Phys. A: Math. Gen.}}}

\author{G. PAGNINI$^1$ and R.K. SAXENA$^2$\\
\\
$^1$ENEA, Centre ``Ezio Clementel'',\\
via Martiri di Monte Sole 4, I-40129 Bologna, Italy\\
gianni.pagnini@bologna.enea.it\\
\\
$^2$Department of Mathematics and Statistics,\\
Jan Narain Vyas University, Jodhpur 342005, India\\
ram.saxena@yahoo.com}

\begin{document}
\maketitle

\begin{abstract}
A Voigt profile function emerges in several physical investigations 
(e.g. atmospheric radiative transfer, astrophysical spectroscopy, 
plasma waves and acoustics) 
and it turns out to be the convolution of the Gaussian 
and the Lorentzian densities.
Its relation with a number of special functions 
has been widely derived in literature starting from its
Fourier type integral representation.
The main aim of the present paper is to introduce the Mellin-Barnes
integral representation as a useful tool
to obtain new analytical results.
Here, starting from the Mellin-Barnes integral representation,
the Voigt function is expressed in terms of the Fox H-function 
which includes representations in terms of the Meijer G-function and previously 
well-known representations with other special functions.
\end{abstract}


\section{Introduction}
During the recent past, representations of the Voigt profile function in terms of
special functions have been discussed in this {\it Journal}
\cite{exton-jpa-1981,katriel-jpa-1982,fettis-jpa-1983,keshavamurthy-jpa-1987,sajo-jpa-1993,
gubner-jpa-1994}.
With the present paper we aim to continue those researches.
In particular, we derive 
the representation in terms of Fox H-function, which includes previously well-known
results.

The Voigt profile function emerges in several physical investigations as
atmospheric radiative transfer, astrophysical spectroscopy, plasma waves and acoustics and
molecular spectroscopy in general. 
Mathematically, it turns out to be the convolution of the Gaussian
and the Lorentzian densities. Here we are studying the ordinary Voigt function
and not its mathematical generalizations, for example,
see the papers \cite{sampoorna_etal-jqsrt-2007,srivastava_etal-ass-1987,srivastava_etal-ass-1992}.
The computation of the Voigt profile is an old issue in literature and 
many efforts are directed to evaluate this function with different techniques.
In fact, an analytical explicit representation in terms of elementary functions
does not exist 
and it can be considered a special function itself.
Moreover it is strictly related to the plasma dispersion function 
\cite{fried-conte-1961} and to
a number of special functions as, for example, the confluent hypergeometric function,
the complex complementary error function, 
the Dawson function,
the parabolic cylinder function and the Whittaker function, see e.g. 
\cite{exton-jpa-1981,katriel-jpa-1982,fettis-jpa-1983,
armstrong-jqsrt-1967,schreier-jqsrt-1992,shippony_etal-jqsrt-1993,yang-ijmest-1994}.
All previous representations are derived starting from the integral formula due
to Reiche in 1913 \cite{reiche-1913} that is actually
a Fourier type integral.

The Voigt profile function remains nowadays a mathematically and computationally
interesting problem because computing profiles with high accuracy 
is still an expensive task.
The actual interest on this topic is proven
by several recent papers (2007) on mathematical
\cite{dulov_etal-jqsrt-2007,he_etal-joa-2007,petrov-jqsrt-2007,sampoorna_etal-jqsrt-2007,
srivastava_etal-rjmp-2007,zaghloul-mnras-2007}
and numerical aspects
\cite{leiweke_etal-jqsrt-2007,letchworth_etal-jqsrt-2007,mendenhall-jqsrt-2007}.
A huge collection exists of published works on this topic, but instead to report it,
we give the significative datum
that searching in Google Scholar the strings
``voigt profile function'' and ``voigt function'' the number of files found is $\sim24,600$
and $\sim70,100$, respectively. 

The Mellin-Barnes integrals are a family of integrals in the complex plane
whose integrand is given by the ratio of products of Gamma functions.
Despite of the name, the Mellin-Barnes integrals were initially studied
in 1888 by the Italian mathematician S.~Pincherle 
\cite{pincherle,mainardi_etal-jcam-2003} in a couple of papers on the
duality principle between linear differential equations and linear difference
equations with rational coefficients.
The Mellin-Barnes integrals are strongly related with the 
Mellin transform, in particular with the inverse transformation. 
As shown by O.I. Marichev \cite{marichev-1982}, 
the problem to evaluate integrals can be successfully faced 
with a powerful method mainly based on their reduction
to functions whose Mellin transform is the ratio of product of
Gamma functions and then, after the inversion, 
the problem consists in the evaluation of Mellin-Barnes integrals.
Moreover, they are also the essential tools for treating
higher transcendental functions as Fox~H-function and Meijer G-function and
a useful representation to compute asymptotic behaviour of functions
\cite{paris-kaminski}.

The main object of the present paper is to introduce the Mellin-Barnes
integral representation as a useful tool
to obtain new analytical results that in the future can lead to efficient numerical
algorithms for the Voigt function. 
A successful application of such approach has been shown
in \cite{pagnini_etal-granada}, where the parametric evolution equation of
the Voigt function (and its probabilistic generalization) is derived and the 
scaling laws, with respect to the parameter, in the asymptotic regimes
are computed using the Mellin-Barnes integral representation.
After all, this work can be seen also as 
an interesting exercise {\it per se} of application of 
the Mellin-Barnes integral method.

The rest of the paper is organized as follows.
In section 2 the basic definition of the Voigt profile function is given and some {\it classical}
and recent representations are reviewed. In section 3 the Mellin-Barnes integral
representation of the Voigt function is derived and, in section 4
starting from this result, first the Voigt function is expressed in terms of 
the Fox H-function and later, in cascade, the representation with the 
Meijer G-function and other special functions are obtained.
Finally in section 5 the summary and conclusions are given. 

\section{The Voigt profile function}
\subsection{Basic definition}
\begin{figure}
\begin{center}
\includegraphics[width=10cm,angle=-90]{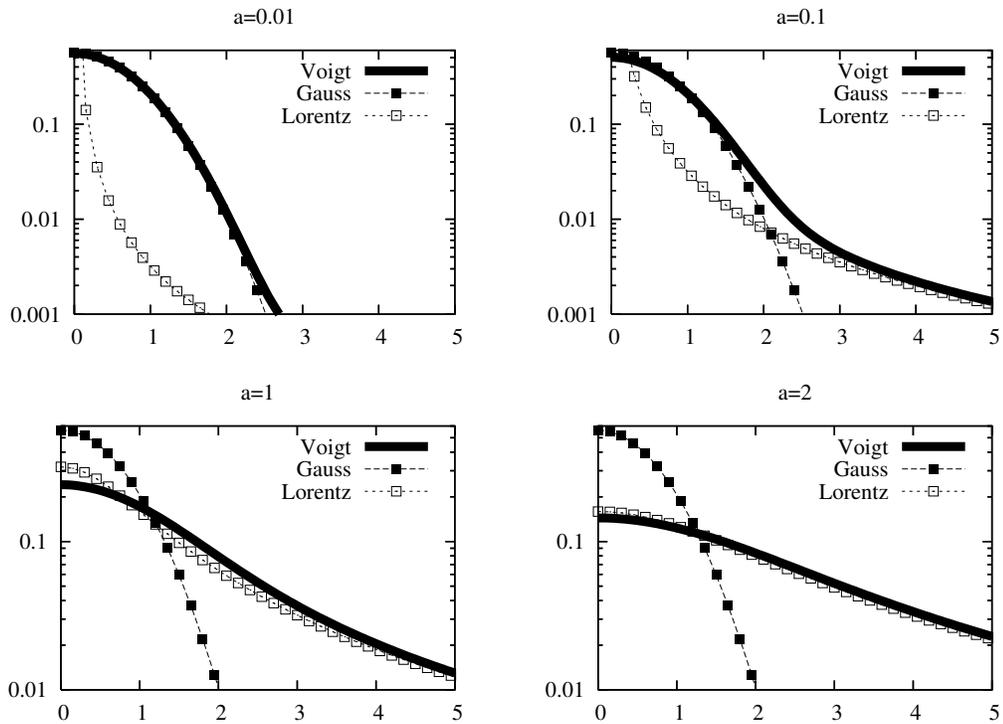}
\end{center}
\caption{Comparison between the Voigtian, Gaussian and Lorentzian profiles with
$a=\omega_L/\omega_G=0.01, 0.1, 1, 2$.}
\label{figura1}
\end{figure}

The Gaussian $G(x)$ and the Lorentzian $L(x)$ profiles 
are defined as

\be
G(x)=\frac{1}{\sqrt{\pi} \omega_G} \, \exp\left[-
\left(\frac{x}{\omega_G}\right)^2 \right] \,,
\quad 
L(x)=\frac{1}{\pi \omega_L} \, \frac{\omega_L^2}{x^2 + \omega_L^2} \,,
\label{g-l}
\ee
where $\omega_G$ and $\omega_L$ are the corresponding widths.
The variable $x$ is a wave-number and then its physical dimension is
a length raised up to $-1$.
The Voigt profile $V(x)$ is obtained by the convolution of $G(x)$ and $L(x)$

\be
V(x)=\int_{-\infty}^{+\infty} L(x-\xi) G(\xi) \, d\xi = 
\frac{\omega_L/\omega_G}{\pi^{3/2}}
\int_{-\infty}^{+\infty}
\frac{\e^{-(\xi/\omega_G)^2}}
{(x-\xi)^2 + \omega_L^2} \, d\xi \,.
\label{voigt}
\ee
The comparison between the Voigtian, Gaussian and Lorentzian profile is shown
in figure~\ref{figura1} for different values of the width ratio $\omega_L/\omega_G$.

Let $\widehat{f}(\k)$ be the Fourier transform of $f(x)$ so that

\be
\widehat{f}(\k)=\int_{-\infty}^{+\infty} \e^{+i \k x} f(x) \, d x \,,
\quad 
f(x)=\frac{1}{2\pi} \int_{-\infty}^{+\infty} \e^{-i \k x} \widehat{f}(\k) \, d\k \,,
\label{fourier}
\ee
then

\be
\widehat{V}(\k)=\widehat{G}(\k) \widehat{L}(\k)= 
\e^{-\omega_G^2 \k^2/4} \,
\e^{-\omega_L|\k|} \,,
\label{fourier-V}
\ee
and

\begin{eqnarray}
V(x) &=& \frac{1}{2\pi} \int_{-\infty}^{+\infty}	
\e^{-i \k x} \e^{-\omega_G^2 \k^2 /4 - \omega_L |\k|} \, d \k \nonumber \\
&=& \frac{1}{\pi} \int_0^{+\infty} 
\e^{-\omega_L \k - \omega_G^2 \k^2/4} \cos(\k x) d \k \,.
\label{reiche}
\end{eqnarray}
Formula (\ref{reiche}) is the integral representation 
due to Reiche \cite{reiche-1913}.

\subsection{Some {\it classical} and recent representations}
\label{representations}
Let $x'$ be the dimensionless variable $x'=x/\omega_G$, 
the Voigt function can be re-arranged in the form

\be
V(x)=\frac{1}{\sqrt{\pi}\omega_G} K(x',y) \,,
\quad
K(x,y)=\frac{y}{\pi}
\int_{-\infty}^{+\infty}
\frac{\e^{-\xi^2}}
{(x-\xi)^2 + y^2} d\xi \,,
\label{H}
\ee
where $y=\omega_L/\omega_G$ and from (\ref{fourier}) it follows that

\be
K(x,y)=\frac{1}{\sqrt{\pi}} \, \int_0^{+\infty}
\e^{-y \xi - \xi^2/4} \cos(x \xi) \, d \xi \,.
\label{alter1}
\ee
The Voigt function does not possess an explicit representation in terms of
elementary functions and several 
alternatives to (\ref{voigt}) have been given in literature, mainly with the
intention to obtain a more efficient numerical computation. 

Combining $x$ and $y$ in the complex variable $z=x-iy$, the function
$K(x,y)$ (\ref{H}) is 

\be
K(x,y)=Re[W(z)] \,, \quad
W(z)=\frac{i}{\pi}\int_{-\infty}^{+\infty}
\frac{\e^{-\xi^2}}{z-\xi} \, d \xi \,.
\label{alter2}
\ee
where $W(z)$ is strongly related to the plasma dispersion function
\cite{fried-conte-1961} and some {\it classical} representations can be found,
see e.g. \cite{armstrong-jqsrt-1967,schreier-jqsrt-1992,shippony_etal-jqsrt-1993}.
In fact,  
form the relation of $W(z)$ with the complex complementary error function $Erfc(-iz)$
and the Dawson function $F(z)=e^{-z^2} \int_0^z \e^{\xi^2} d\xi$ then it follows that

\be
K(x,y)=Re[W(z)] \,, \quad W(z)=e^{-z^2} Erfc(-iz) \,,
\quad y>0 \,,
\label{alter3}
\ee

\be
K(x,y)=Re[W(z)] \,, \quad W(z)=e^{-z^2} + \frac{2i}{\sqrt{\pi}} F(z) \,.
\label{alter4}
\ee

More recent representations are, for example, those derived in 2001 by 
Di Rocco {\it et al} \cite{dirocco_etal-as-2001}

\begin{eqnarray}
K(x,y)=\sum_{n=0}^\infty (-1)^n \left\{
\frac{1}{\Gamma(n+1)} {_1}F_1 
\left(\frac{2n+1}{2},\frac{1}{2}; y^2\right) - \right. \nonumber \\
\qquad \qquad \qquad \qquad \qquad 
\left. \frac{2a}{\Gamma\left(\frac{2n+1}{2}\right)} 
{_1}F_1 \left(n+1,\frac{3}{2}; y^2\right) \right\} x^{2n} \,,
\label{alter5}
\end{eqnarray}
where $_1F_1(\alpha,\beta;z)$ is the confluent hypergeometric function,
and in 2007 by Zaghloul \cite{zaghloul-mnras-2007}, which ``completed'' a previous
formula given by Roston \& Obaid \cite{roston_etal-jqsrt-2005},

\be
K(x,y) = [1-{\erf}(y)] \, e^{(-x^2+y^2)} \, {\cos}(2xy) + 
\frac{2}{\sqrt{\pi}} 
\int_0^x \e^{(-x^2+\xi^2)} \sin[2y(x-\xi)] \, d\xi \,.
\label{alter6}
\ee 

Recently (2007), He \& Zhang \cite{he_etal-joa-2007}
claimed to have derived an exact calculation of the Voigt profile
that is proportional to the product of an exponential and a cosine function.
However this representation assumes negative values in contrast with the non-negativity
of the Voigt function. For this reason, this result is not correct.

Further representations are given in terms of special functions, see for example
the one involving the confluent hypergeometric function $_1F_1$ 
\cite{katriel-jpa-1982,fettis-jpa-1983}

\begin{eqnarray}
K(x,y)=e^{(y^2-x^2)} {\cos}(2xy) -
\frac{1}{\sqrt{\pi}} \left\{
(y+ix) \, _1F_1(1;3/2;(y+ix)^2) \, + \right. \nonumber \\
\qquad \qquad \qquad \qquad
\qquad \qquad
\left. (y-ix) \, _1F_1(1;3/2;(y-ix)^2) \right\}
\,,
\label{F}
\end{eqnarray}
and others involving the Whittaker function $W_{k,m}$,
the $Erfc-$function and the parabolic cylinder function 
\cite[formulae (17,13,16)]{yang-ijmest-1994}

\begin{eqnarray}
K(x,y)=\frac{1}{2 \sqrt{\pi}} \left\{
(y-ix)^{-1/2} \, e^{(y-ix)^2/2} \, W_{-1/4,-1/4}((y-ix)^2) \, + \right. \nonumber \\
\qquad \qquad \qquad \qquad
\left. (y+ix)^{-1/2} \, e^{(y+ix)^2/2} \, W_{-1/4,-1/4}((y+ix)^2)
\right\} \,,
\label{W}
\end{eqnarray}

\be
K(x,y)=\frac{1}{2} \left\{
e^{(y-ix)^2} \, Erfc(y-ix) + e^{(y+ix)^2} \, Erfc(y+ix)
\right\} \,,
\label{E}
\ee

\be
K(x,y)=\frac{e^{(y^2-x^2)/2}}{\sqrt{2 \pi}} \left\{
e^{-ixy} \, D_{-1}[\sqrt{2}(y-ix)] + e^{+ixy} \, D_{-1}[\sqrt{2}(y+ix)]
\right\} \,.
\label{P}
\ee

\section{The Mellin-Barnes integral representation}
Let us consider again dimensional variables and the Gauss, Lorentz and Voigt functions
defined as in (\ref{g-l}-\ref{voigt}). 
From (\ref{fourier-V}) we have

\be
V(x) = \frac{1}{2\pi} \int_{-\infty}^{+\infty}	
\e^{-i \k x} \e^{-\omega_G^2 \k^2 /4 - \omega_L |\k|} \, d \k 
= \frac{1}{2\pi} \left\{I_+(x) + I_-(x) \right\} \,,
\ee
where

$$
I_{\pm}(x)= \int_0^{+\infty} e^{-(\omega_L \pm ix)\k - \omega_G^2 \k^2 /4} \, d \k \,.
$$
Following \cite{paris-kaminski}, 
the Mellin-Barnes integral representations of $I_\pm(x)$ can be obtained 
from the definition of the Gamma function 

$$
\Gamma(z)=\int_0^{+\infty} \xi^{z-1} \e^{-\xi} \, d \xi \,,
$$
and the Mellin-Barnes integral representation of $\e^{-z}$ $(z \ne 0)$

$$
e^{-z}=\frac{1}{2\pi i} \int_\mathcal{L} \Gamma(s) z^{-s} ds =
\sum_{n=0}^{\infty} \frac{(-1)^n}{n!} z^n \,,
$$
where $\mathcal{L}$ denotes a loop in the complex $s$ plane which encircles the poles
of $\Gamma(s)$ (in the positive sense) with endpoints at infinity in $Re(s) < 0$
and with no restrictions on $\arg z$ \cite{paris-kaminski}.
The functions $I_\pm(x)$ have the following Mellin-Barnes integral representations

\begin{eqnarray}
I_\pm (x) &=& \int_0^{+\infty} \e^{-\omega_G^2 \k^2 /4} 
\left\{\frac{1}{2\pi i} \int_{\mathcal{L}} \Gamma(s) 
[(\omega_L \pm ix)\k]^{-s} \, ds \right\} d \k \nonumber \\
&=& \frac{1}{2\pi i} \int_{\mathcal{L}} \Gamma(s) 
\left\{\int_0^{+\infty} \e^{-\omega_G^2 \k^2 /4} \k^{-s} \, ds \right\}
(\omega_L \pm ix)^{-s} \, d\k \nonumber \\
&=& \frac{1}{\omega_G} 
\frac{1}{2\pi i} \int_{\mathcal{L}} \Gamma(s) \Gamma\left(\frac{1}{2}-\frac{s}{2}\right)
\left[\frac{2}{\omega_G}(\omega_L \pm ix)\right] ^{-s} \, ds \,.
\label{MB0}
\end{eqnarray}
Hence the Mellin-Barnes integral representation of the Voigt function is
given by

\begin{eqnarray}
V(x)=\frac{1}{2 \pi \omega_G} \left\{
\frac{1}{2\pi i} \int_{\mathcal{L}} \Gamma(s) \Gamma\left(\frac{1}{2}-\frac{s}{2}\right)
\left[\frac{2}{\omega_G}(\omega_L + ix)\right] ^{-s} \, ds \right. + \nonumber \\
\qquad \qquad \qquad \qquad
\left.
\frac{1}{2\pi i} \int_{\mathcal{L}} \Gamma(s) \Gamma\left(\frac{1}{2}-\frac{s}{2}\right)
\left[\frac{2}{\omega_G}(\omega_L -  ix)\right] ^{-s} \, ds \right\}\,.
\label{MB1}
\end{eqnarray}
However we note that for a complex number $z=|z| e^{i\theta}$, $\theta=\arctan(Im(z)/Re(z))$,
the following rule holds

\begin{eqnarray*}
z^n + \bar{z}^n &=& |z|^n e^{i n \theta} + |z|^n e^{-i n \theta} \\
&=& |z|^n (e^{i n \theta} + e^{-i n \theta})\\
&=& 2 |z|^n \cos(n\theta) = 2 |z|^n \cos(n \arctan(Im(z)/Re(z))) \,,
\end{eqnarray*}
and formula (\ref{MB1}) becomes

\be
V(x)=
\frac{1}{\pi \omega_G} \,
\frac{1}{2\pi i} \int_{\mathcal{L}} \Gamma(s) \Gamma\left(\frac{1}{2}-\frac{s}{2}\right)
\cos\left[s \arctan\left(\frac{x}{\omega_L}\right)\right]
\left(4\frac{\omega_L^2+x^2}{\omega_G^2}\right)^{-s/2} ds \,.
\label{MB1-bis}
\ee

Consider again (\ref{MB0}), 
changing $s \to -s$ and taking the corresponding integration path $\mathcal{L}$
as a loop in the complex plane that encircles the poles of $\Gamma(-s)$,
an other Mellin-Barnes integral representation of the Voigt function equivalent to (\ref{MB1}) is

\begin{eqnarray}
V(x)=\frac{1}{2\pi \omega_G} \left\{
\frac{1}{2\pi i} \int_{\mathcal{L}} \Gamma(-s) \Gamma\left(\frac{1}{2}+\frac{s}{2}\right)
\left[\frac{2}{\omega_G}(\omega_L + ix)\right] ^s \, ds \right. \, + \nonumber \\
\qquad \qquad \qquad \qquad
\left.
\frac{1}{2\pi i} \int_{\mathcal{L}} \Gamma(-s) \Gamma\left(\frac{1}{2}+\frac{s}{2}\right)
\left[\frac{2}{\omega_G}(\omega_L -  ix)\right] ^s \, ds \right\} \,,
\label{MB2}
\end{eqnarray}
and in more compact form 

\be
V(x)=
\frac{1}{\pi \omega_G} \,
\frac{1}{2\pi i} \int_{\mathcal{L}} \Gamma(-s) \Gamma\left(\frac{1}{2}+\frac{s}{2}\right)
\cos\left[s \arctan\left(\frac{x}{\omega_L}\right)\right]
\left(4\frac{\omega_L^2+x^2}{\omega_G^2}\right)^{s/2} ds \,.
\label{MB2-bis}
\ee

\section{The Fox H- and Meijer G-function representations}
\subsection{The Fox H-function representation}
The Voigt function can be represented in terms of 
well known special function, see \S \ref{representations}, but its representation in terms
of Fox H-function is still not known.
The expression in terms of Fox H-function is important because it is 
the most modern representation method and, actually, it is the most compact form to represent
higher transcendental functions. The definition of the Fox H-function is given in Appendix.

For $\omega_G$ and $\omega_L$ fixed, $2(\omega_L+ix)/\omega_G \ne 0$ and
$2(\omega_L-ix)/\omega_G \ne 0$, from (\ref{MB2}) and (\ref{MB1}) we have,
respectively, 

\begin{eqnarray}
V(x)=\frac{1}{2\pi \omega_G} \left\{
H^{11}_{11}\left[
\frac{2}{\omega_G}(\omega_L + ix) \left|
\begin{array}{c}
(1/2,1/2)\\
(0,1)
\end{array} \right. \right] \right. + \nonumber \\ 
\qquad \qquad \qquad \qquad
\left. H^{11}_{11}\left[
\frac{2}{\omega_G}(\omega_L - ix) \left|
\begin{array}{c}
(1/2,1/2)\\
(0,1)
\end{array} \right. \right] \right\} \,, 
\label{H1}
\end{eqnarray}

\begin{eqnarray}
V(x)=\frac{1}{2\pi \omega_G} \left\{
H^{11}_{11}\left[
\frac{\omega_G}{2 (\omega_L + ix)} \left|
\begin{array}{c}
(1,1)\\
(1/2,1/2)
\end{array} \right. \right] \right. + \nonumber \\ 
\qquad \qquad \qquad \qquad
\left. H^{11}_{11}\left[
\frac{\omega_G}{2(\omega_L - ix)} \left|
\begin{array}{c}
(1,1)\\
(1/2,1/2)
\end{array} \right. \right] \right\} \,,
\label{H2}
\end{eqnarray}
As a consequence of the fact that this is the most 
comprehensive representation, in cascade, the others with
less general functions can be obtained.

\subsection{The Meijer G-function representation}
The first H-function in (\ref{H1}) can be rewritten in 
terms of the Meijer G-function. Setting
$Z=2(\omega_L+ix)/\omega_G$, we have

\begin{eqnarray}
H^{11}_{11}\left[ Z \left|
\begin{array}{c}
(1/2,1/2)\\
(0,1)
\end{array} \right. \right] &=&
\frac{1}{2 \pi i}\int_{\mathcal{L}} \Gamma(-s)\Gamma\left(\frac{1}{2}+\frac{s}{2}\right)
Z^s ds \nonumber \\
&=& \frac{2}{2 \pi i}\int_{\mathcal{L}} \Gamma(-2s)\Gamma\left(\frac{1}{2}+s\right)
Z^{2s} ds \nonumber \\ 
&=& \frac{1}{\sqrt{\pi}} \frac{1}{2 \pi i}\int_{\mathcal{L}} 
\Gamma(-s)\Gamma\left(\frac{1}{2}-s\right)\Gamma\left(\frac{1}{2}+s\right)
\left(\frac{Z^2}{4}\right)^s ds \nonumber \\ 
&=& \frac{1}{\sqrt{\pi}} \, G^{21}_{12}\left[\frac{Z^2}{4} \left|
\begin{array}{c}
1/2\\
0,1/2
\end{array} \right. \right] \,,
\end{eqnarray}
where the change of variable $s \to 2s$ and
the duplication rule for the Gamma function,
$\Gamma(2z)=\Gamma(z)\Gamma(1/2 + z) 2^{2z-1} \pi^{-1/2}$, are applied.
The second H-function in (\ref{H1}) follows from the first
with $\bar{Z}=2(\omega_L -ix)/\omega_G$. Finally, the Voigt function
in terms of the Meijer G-function is given by

\be
V(x)=\frac{1}{2 \pi^{3/2} \omega_G} \left\{
G^{21}_{12}\left[
\frac{Z^2}{4} \left|
\begin{array}{c}
1/2\\
0,1/2
\end{array} \right. \right] 
+ G^{21}_{12}\left[
\frac{{\bar{Z}}^2}{4} \left|
\begin{array}{c}
1/2\\
0,1/2
\end{array} \right. \right] \right\} \,. 
\label{G1}
\ee
The Meijer G-function can be reduced to other special functions and,
for example, representations given in (\ref{W}, \ref{E}, \ref{P}) 
are straightforwardly recovered. 

In fact, the Meijer G-function and the Whittaker function $W_{k,m}$ are related
by the formula \cite[p. 435]{erdely_etal-1954}

$$
G^{21}_{12}\left[ z \left|
\begin{array}{c}
a\\
b,c
\end{array} \right. \right] =
\Gamma(b-a+1)\Gamma(c-a+1) x^{(b+c-1)/2} e^{x/2} W_{k,m}(k) \,,
$$
where $k=a-(b+c+1)/2$ and $m=(b-c)/2$. Thus the G-functions in (\ref{G1})
are expressed in terms of the Whittaker function as

$$
G^{21}_{12}\left[ \frac{Z^2}{4} \left|
\begin{array}{c}
1/2\\
0,1/2
\end{array} \right. \right] =
\sqrt{\pi} \, \left(\frac{Z}{2}\right)^{-1/2} \, e^{Z^2/8} \, W_{-1/4,-1/4}(Z^2/4) \,.
$$
Hence, the Voigt function in terms of the Whittaker function is 

\begin{eqnarray}
V(x)=\frac{1}{2 \pi \omega_G} \left\{
\left(\frac{Z}{2}\right)^{-1/2} \, e^{Z^2/8} \, W_{-1/4,-1/4}(Z^2/4) + \right. \nonumber \\
\qquad \qquad \qquad \qquad \qquad
\left. \left(\frac{\bar{Z}}{2}\right)^{-1/2} \, e^{{\bar{Z}}^2/8} \, W_{-1/4,-1/4}({\bar{Z}}^2/4)
\right\} \,.
\label{V-W}
\end{eqnarray}
Moreover, from the identity

$$
G^{21}_{12}\left[ \frac{Z^2}{4} \left|
\begin{array}{c}
1/2\\
0,1/2
\end{array} \right. \right] =
\pi \, e^{Z^2/4} \, Erfc(Z/2) \,,
$$
we have 

\be
V(x)=\frac{1}{2 \sqrt{\pi} \omega_G} \left\{
e^{Z^2/4} \, Erfc(Z/2) + e^{{\bar{Z}}^2/4} \, Erfc({\bar{Z}}/2)
\right\} \,,
\label{V-E}
\ee
and from 

$$
G^{21}_{12}\left[ \frac{Z^2}{4} \left|
\begin{array}{c}
1/2\\
0,1/2
\end{array} \right. \right] =
\sqrt{2 \pi} \, e^{Z^2/8} \, D_{-1}(Z/\sqrt{2}) \,,
$$
we have 

\be
V(x)=\frac{1}{\sqrt{2} \pi \omega_G} \left\{
e^{Z^2/8} \, D_{-1}(Z/\sqrt{2}) + e^{{\bar{Z}}^2/8} \, D_{-1}({\bar{Z}}/\sqrt{2})
\right\} \,.
\label{V-P}
\ee
Setting $\omega_G=1$ in (\ref{V-W},\ref{V-E},\ref{V-P}) 
formulae (\ref{W},\ref{E},\ref{P}) are recovered, respectively.

\section{Summary and conclusions}
In the present paper the Mellin-Barnes integral representation of the
Voigt profile function is derived. We think that this integral representation 
is a useful tool to have new analytical and numerical results in the subject. 
Starting from this, the
Voigt function has been expressed in terms of the Fox H-function, which
is the most comprehensive representation. In cascade, 
the expression in terms of the Meijer G-function is obtained and the previous well-known 
representations with the Whittaker, the $Erfc$ and the Parabolic cylinder functions
are recovered.


\appendix
\section*{Appendix. The Fox H- and the Meijer G- functions}
The Fox H-function is defined as \cite{mathai-saxena,srivastava_etal-1982} 

$$
H^{mn}_{pq} \left[ z \left|
\begin{array}{c}
(a_1,A_1), \cdots, (a_p,A_p) \\
(b_1,B_1), \cdots, (b_q,B_q) 
\end{array} \right. \right] =
\frac{1}{2\pi i} \int_{\mathcal{L}} h(s) z^s ds \,,
$$
where

$$
h(s)=\frac{\prod_{j=1}^m \Gamma(b_j-B_j s) \prod_{j=1}^n \Gamma(1-a_j+A_j s)}
{\prod_{j=m+1}^q \Gamma(1-b_j+B_j s) \prod_{j=n+1}^p \Gamma(a_j-A_j s)} \,,
$$
an empty product is interpreted as unity,
$\{m\,, \, n \,, \,p \,, \, q\}$ are nonnegative integers so that $0 \le m \le q$, $0 \le n \le p$,
$\{A_j \,, \, B_j\}$ are positive numbers and $\{a_j \,, \, b_j\}$ complex numbers.
The H-function is an analytic function of $z$ and makes sense if 
\cite{mathai-saxena,srivastava_etal-1982}

$$
\begin{array}{lr}
i) \quad \forall z \ne 0 \,, & \mu > 0 \\
\\
ii) \quad  0 < |z| < \beta^{-1} \,, & \mu=0
\end{array}
\quad ,
$$
where

$$
\mu=\sum_{j=1}^q B_j -\sum_{j=1}^p A_j \,, \quad 
\beta=\prod_{j=1}^p A_j^{A_j} \prod_{j=1}^q B_j^{-B_j} \,.
$$

The Meijer G-function corresponds to the special case 
$A_j=B_k=1$ $(j=1, \dots,p \,; k=1, \dots,q)$

$$
H^{mn}_{pq} \left[ z \left|
\begin{array}{c}
(a_1,1), \cdots, (a_p,1) \\
(b_1,1), \cdots, (b_q,1) 
\end{array} \right. \right] =
G^{mn}_{pq} \left[ z \left|
\begin{array}{c}
a_1, \cdots, a_p \\
b_1, \cdots, b_q 
\end{array} \right. \right] \,.
\label{G}
$$


\end{document}